# 3D Texture Coordinates on Polygon Mesh Sequences

Eric Mootz, June 2014

*Cecilienstraße 15, D-66111 Saarbrücken, Germany*

*info@mootzoid.com*

***Abstract*— A method for creating 3D texture coordinates for a sequence of polygon meshes with changing topology and vertex motion vectors.**

## I. Introduction

The use of polygon mesh sequences as a final form for animated implicit surfaces presents the challenge of generating meaningful and temporally coherent texture coordinates.

An easily implemented method for creating 3D texture coordinates is proposed and presented as a technique relevant for meshing algorithms as for example the widely used marching cubes [1]. In general these sequences have changing topology: the amount of points (vertices) and polygons varies between each element of the sequence.

The texture coordinates are generated after the mesh and it is therefore irrelevant with what method the meshes are created.

The proposed method has been implemented with success in the two commercial plugins *emPolygonizer* and *emTopolizer*.

## II. Formal Conventions

In the following *3D texture coordinates* will be referred to as UVWs and a *sequence of polygon meshes* will be referred to as M, the amount of elements (polygon meshes) in M as $|M|$ and its elements as $M_i$ with $1 \leq i \leq |M|$.

Each element $M_i$ in M is described by a *set of vertices* $V_{M_i}$ and a *polygonal description* $P_{M_i}$ that contains the triangles as triplets of vertex indices. $|V_{M_i}|$ is the amount of vertices in $V_{M_i}$ and $V_{M_{i,j}}$ the j-th vertex with $1 \leq j \leq |V_{M_i}|$. $|P_{M_i}|$ is the amount of polygons (triangles) in $P_{M_i}$ and $P_{M_{i,j}}$ the j-th triangle with $1 \leq j \leq |P_{M_i}|$. The vertex indices of the j-th triangle are $P_{M_{i,j,1}}$, $P_{M_{i,j,2}}$ and $P_{M_{i,j,3}}$.

Each vertex $V_{M_{i,j}}$ has a *position* $V^{Pos}_{M_{i,j}}$, a *motion vector* $V^{Vel}_{M_{i,j}}$ and a *texture coordinate* $V^{Tex}_{M_{i,j}}$, all of which are in 3D space.

A *location* L on $M_i$ is defined as the barycentric coordinates of a triangle in $P_{M_i}$.

## III. Method

For a given sequence M, initialize the texture coordinates of $M_1$, then, for each $M_i$ in M, backtrace the vertex positions $V^{Pos}_{Mi}$ using the motion vectors $V^{Vel}_{Mi}$ and transfer the texture coordinates of the closest locations on $M_{i-1}$ on to $M_i$, with $i \geq 2$.

Pseudo code:

```
Initialize V^Tex_M₁
For i = 2 to |M|
    For j = 1 to |V_Mi|
        p = V^Pos_Mi,j - V^Vel_Mi,j
        L = Closest Location on M_i-1 to p
        V^Tex_Mi,j = Texture Coordinate of L
    Next j
Next i
```

## IV. Results

The two commercial plugins *emPolygonizer (http://www.mootzoid.com/plugin/empolygonizer/latest)* and *emTopolizer (http://www.mootzoid.com/plugin/emtopolizer/latest)* have both been using the presented method since end of 2012 to create UVWs on mesh sequences, simulated as well as cached.

## V. Discussion

The algorithm presented can be implemented either in existing code or in a node based 3D application such as Softimage's ICE or Houdini, both of which provide the necessary functionality to find the closest location on a polygon mesh.

As with most simulations the initial state (i.e. the initial UVWs) is crucial and will determine the main look of the result.

The method can be used on existing, i.e. cached, mesh sequences or directly during mesh generation by keeping a polygon mesh $M_n$ in memory and then using it for the creation of the texture coordinates of the next mesh $M_{n+1}$.

The UVWs should have at least 32 bit precision (float) in order to produce acceptable results over time. Tests with 16 bit half float UVWs resulted in too many rounding errors that got worse the longer the sequence was, each error being passed on to the next mesh in the sequence.

Polygon meshes tend to have slightly jittering UVWs with this method, especially on coarse meshes. On high resolution meshes the jittering is hardly noticeable and can furthermore be reduced by applying a Laplacian Smooth to the UVW values.

Some tests revealed a minor tendency of the UVWs to drag behind (or hurry ahead). One reason for that behavior are little imprecisions in the calculations of the backward advection. Another reason is the nature of the motion vectors itself: they can either represent the *current motion* (i.e. the motion that defines the vertex position for the next frame) or the *previous motion* (i.e. the motion that was used in the previous frame to get to the current vertex position), depending on how the algorithm that creates the mesh was implemented by the programmer. The drag/hurry effect can be reduced by using a sort of backward-forward advection together with an error compensation, similar to the BFECC method [2] [3] used in grid based fluid solvers [4].

## VI. Conclusions

The present method produces quite satisfactory results and is easy to implement, especially when using a preexisting library or software for the gathering of the closest locations.

The method has been implemented in commercial plugins such as *emPolygonizer* (version 4.0 and above) and *emTopolizer* (version 1.0 and above) and used in various commercial productions.

## Acknowledgment

Thanks to Andy Moorer, Shannon Moon, Dr. Anne S. Dederichs and Sandra E. Wildt for their time and advice.

## References

[1]   William E. Lorensen and Harvey E. Cline, *Marching Cubes: a high resolution 3D surface construction algorithm*, Computer Graphics, Volume 21, Number 4, July 1987.

[2]   R. MacCormack. *The effect of viscosity in hypervelocity impact cratering*. In AIAA Hypervelocity Impact Conference, 1969. AIAA paper 69-354.

[3]   Selle, A., Fedkiw, R., Kim, B., Liu, Y., and RossignacC, J. 2008. *An unconditionally stable MacCormack method*. J. Sci. Comp. 35, 350–371.

[4]   Stam, J. 1999. *Stable fluids*. In Proceedings of ACM SIGGRAPH, 121–128.

## Figures

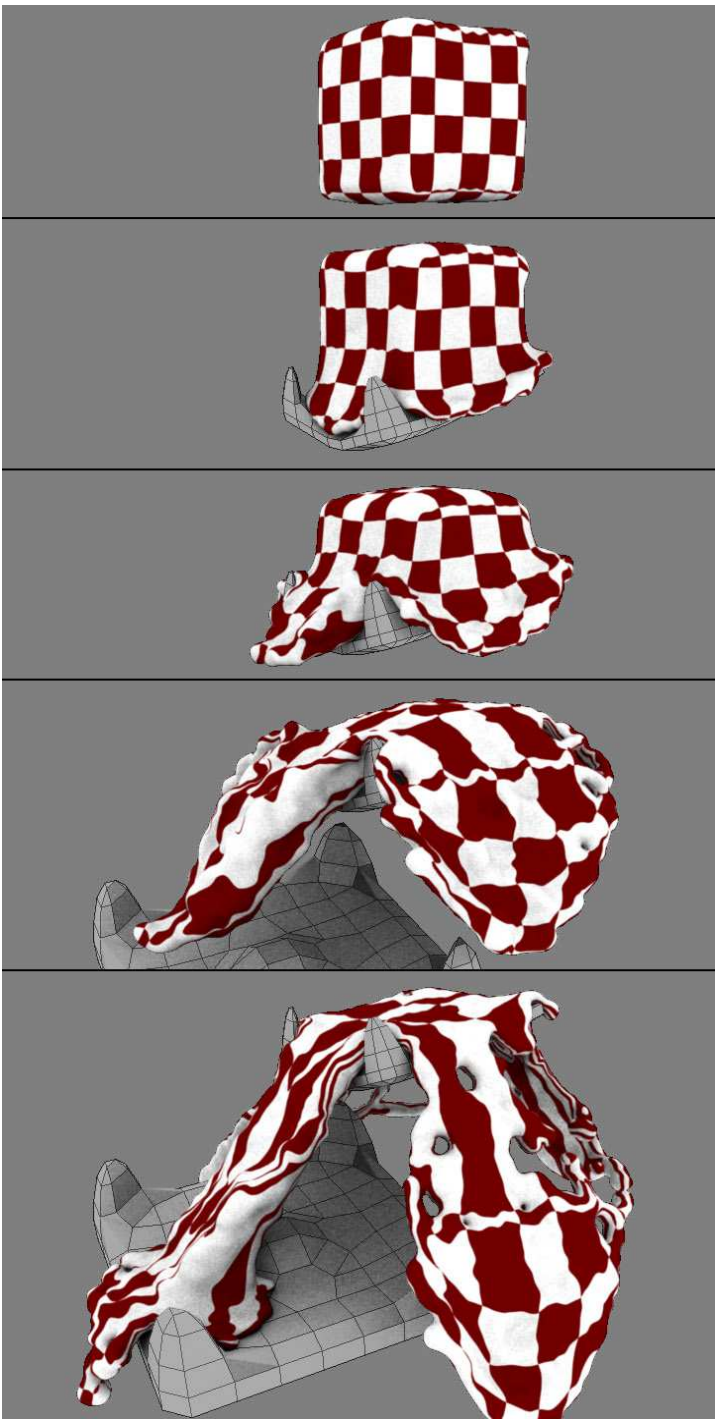 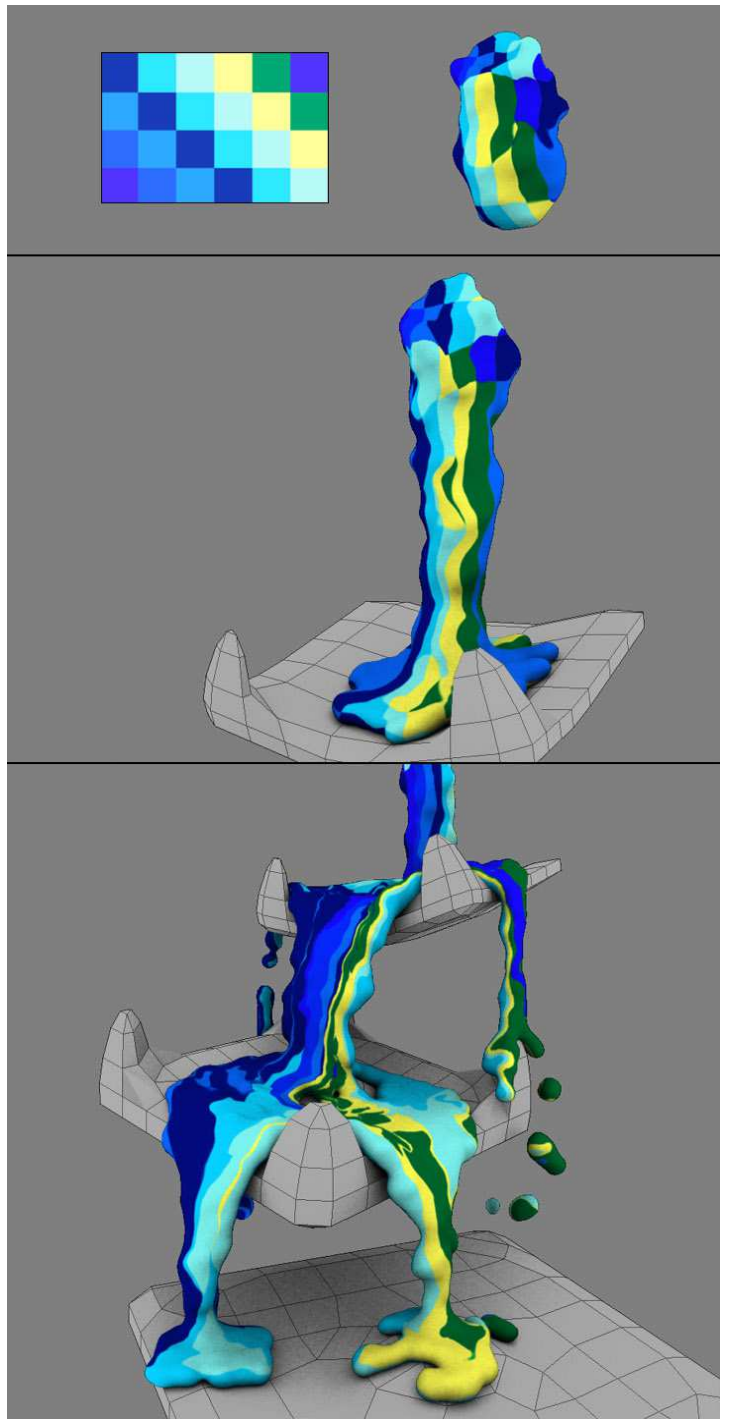

Fig. 1 Texture coordinates generated while meshing a point cloud with the plugin *emPolygonizer4* for Softimage and Maya. The renderings show the final meshes with a procedural 3D checkerboard.

Fig. 2 UVWs generated for a sequence of geometry cache files using the *UVW Engineer* of the plugin *emTopolizer* for Softimage's ICE. The renderings show the final meshes with a 2D image texture.

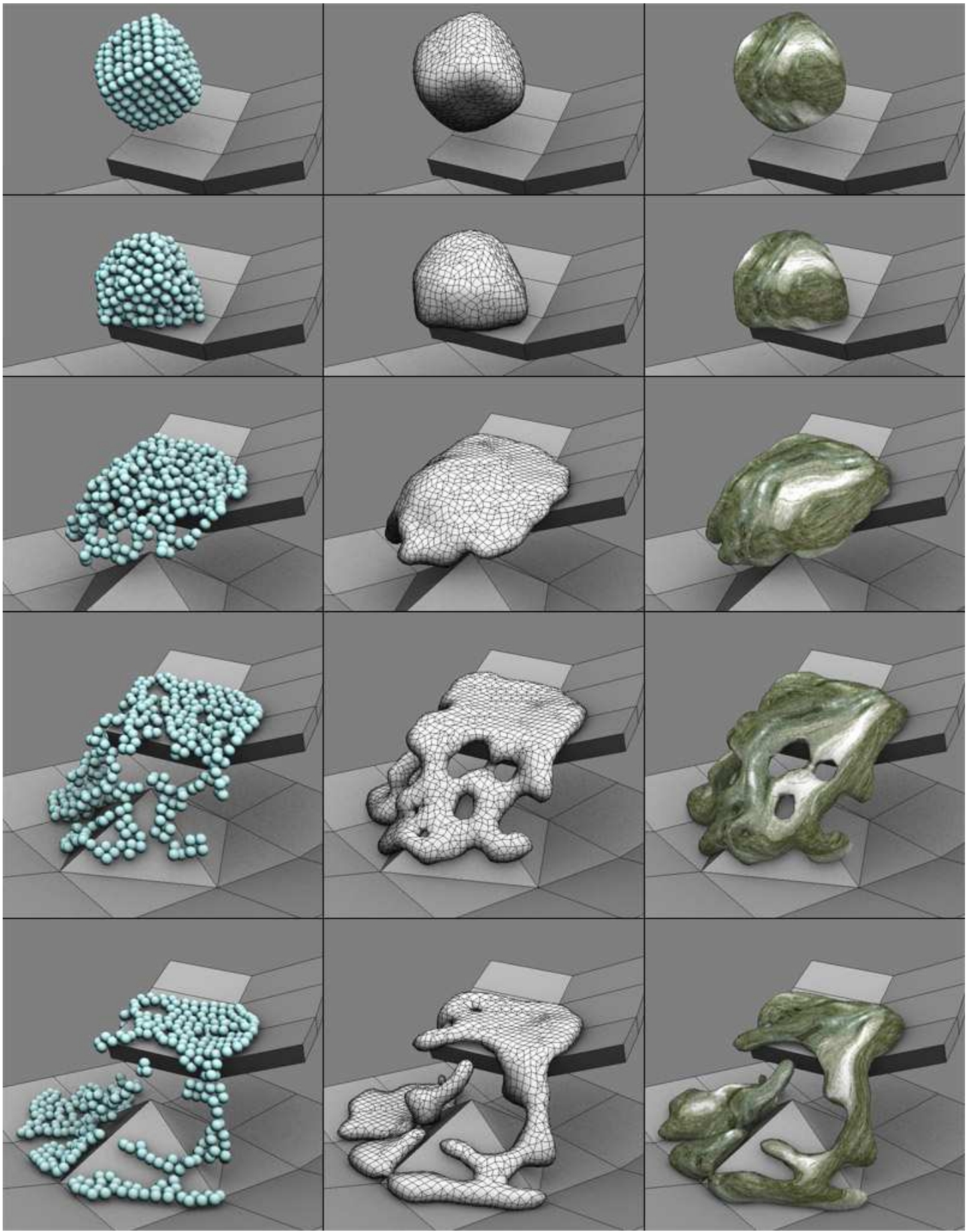

Fig. 3 Left: the particles of a simple SPH liquid simulation. Middle: the meshed particles using marching cubes and some post smoothing and quadrangulation. The UVWs are generated directly afterwards using the method of this paper. Right: the rendered mesh with a texture using the generated UVWs.